\documentclass{article}
\usepackage{latexsym}
\usepackage{stmaryrd}
\usepackage{amssymb,amsmath,mathrsfs,a4wide}
\usepackage{url}

\newcommand{\ourtool}{{\tt G-ACM}~}

\newcommand{\grant}{{\sf grant}}
\newcommand{\deny}{{\sf deny}}

\newcommand{\type} {{\sf type}}

\usepackage{color}
\usepackage{graphicx}

\newtheorem{definition}{Definition}

\title{The \ourtool Tool: using the Drools Rule Engine for Access Control Management}
\author{Jo\~ao S\'a \and Sandra Alves$^{1}$\thanks{Partially funded by FCT, Portuguese Foundation for Science and Technology within project UID/EEA/50014/2013.} \and Sabine Brodab$^{2}$\thanks{Partially supported by CMUP (UID/MAT/00144/2013), which is funded by FCT (Portugal) with national (MEC) and European structural funds through the programs FEDER, under the partnership agreement PT2020.}\\\\
  $^1$ CRACS/INESCTEC, Faculty of Sciences, University of Porto\\
  $^2$ CMUP, Faculty of Sciences, University of Porto
 }
\begin{document}
\maketitle

\begin{abstract}
In this paper we explore the usage of rule engines in a graphical framework for visualising dynamic access control policies. We use the Drools rule engine to dynamically compute permissions, following the Category-Based Access Control metamodel.

\end{abstract}

\section{Introduction}
With the urge to efficiently protect resources from unauthorised access and to preserve the confidentiality of critical data, a wide range of access control models have been developed  (such as MAC, the ANSI H-RBAC, DEBAC, etc.~\cite{bell76,sandhu96,Barker-etal:07}). Despite the variety of models, with each model developing its specific languages, implementation techniques, reasoning methods, properties, etc,  there is a lack of effective methods and tools that facilitate the tasks of policy specification and analysis. Formal specifications of access control models and policies have used theorem provers, purpose-built logics, functional approaches, etc. However, although textual languages and logic-based models are convenient for theoreticians or computer experts, graphical frameworks are more appealing to less technical users, urging the development of models that establish a bridge between theoretical tools and the practitioners' needs. 

In~\cite{AlvesF15},  a framework was presented with the aim of aiding on the specification and analysis of access control policies, based on a metamodel for access control (CBAC) proposed by Barker in~\cite{Barker09}, and that uses a graph-based formalism to represent policies. The CBAC metamodel identifies a core set of principles of access control, abstracting away many of the complexities found in specific models. The expressiveness of the metamodel was shown by deriving well-known access control models, as specific instances of the metamodel~\cite{BertolissiF14}. A key aspect of the metamodel is the notion of \emph{category}, which is a class of entities that share some property.  Classic types of groupings like a role, a security clearance, a discrete measure of trust, etc., are particular instances of the notion of category. In category-based access control policies, permissions are assigned to categories of users, rather than to individual users. 
Categories can be defined as a dynamic propertiy, causing permissions to change autonomously unlike, e.g., role-based access control models, which require the intervention of an administrator.  
A long-term goal was to define a user-friendly language to describe the dynamic behaviour of categories, suitable for policy administrators and that could be translated into other programming languages for integration with policy analysis tools. Security policies are generally written by less technical users, and one major challenge in the specification of policies is the translation of policies from a non-formal specification to a formal one, which can be verified using automated tools. This is a challenging task, due to the complexity of natural language processing, but small steps can be taken in that direction. 

In this paper we present the \ourtool (short for Graphical Access Control Management) tool for the visualisation and analysis of access control policies, inspired by the framework in~\cite{AlvesF15}, and where we use the JBoss Drools rule engine to capture the dynamic behaviour of categories. Our goal is to use Drools to compute permissions based on a set of pre-defined rules that can be changed by the user administrating the policy. In particular, the definitions of categories, therefore the permissions, can be changed by redefining the rules based on the available attributes. 
There is a separation in the graphical framework~\cite{AlvesF15}, between the graphical representation of the policy at a given moment, which corresponds to a static graph where one can use static analysis on graphs to extract properties, and the rewrite-based operational semantics that allows for the permissions to change dynamically. A graph policy can be seen as a snapshot of the system at a particular time from which one can extract the permission of the system at that time, but not the next stage of the graph, which is given by the operational semantics. In our tool this separation is also clear, where Drools is used as a mechanism to derive new instances of the graph, when changes occur in the attributes defining categories. Drools is based on forward-chaining, using actions over facts to infer new facts.
 There are clear advantages in the use of a rule engine in this context: it allows for a separation between the business logic and the application (suitable in this scenario); it simplifies the task of adding new or changing existing customizable facts; it is reasonably understood by less technical users, because of its high-level and declarative syntax, providing a clear correspondence with the CBAC model; efficiency and scalability. 
\section{Related Work} \label{sec:related}

This work is based on the representation of CBAC policies introduced in \cite{AlvesF15}. 
Within CBAC, only textual languages have been used and have focused mainly on the expressivity of the model, the analysis of policies, and the techniques that can be used to enforce policies \cite{Barker09, BertolissiC:essos2010, BertolissiC:STM2010, BertolissiF14, AliF14}.

Several other access control models have been studied through the use of graph-based languages. For example, Koch and al. \cite{KochMP02,KochMP05} use directed graphs to formalize RBAC. A distinctive feature in this work is the use graph transformation rules to model role management operations.
The graphs in \cite{KochMP02,KochMP05} and \cite{AlvesF15} are both typed and labelled. The typing system is similar in both cases but the label structure in \cite{AlvesF15} is richer so it can express policies where access rights depend on data associated to the entities in the policy.  Labels in \cite{KochMP02,KochMP05} are simply identifiers used to encode RBAC.

The RBAC policies in \cite{KochMP02,KochMP05} can be represented by graphs in \cite{AlvesF15}, since a role is a particular case of a category. However, graphs used in \cite{KochMP02,KochMP05} represent also session information, which is not dealt by policy graphs in \cite{AlvesF15}. Nevertheless, since the notion of session in RBAC is similar to the same notion in CBAC, the representation of sessions provided in \cite{KochMP02,KochMP05} could be easily adapted to policy graphs representing CBAC policies.

LaSCO \cite{Hoagland2000} represents policies through graphs in which nodes represent system objects and edges represent system events. In LaSCO, each policy graph represents both the situation under which a policy applies and the constraint that must hold for the policy to be upheld. The model proposed by Koch and al., LaSCO and other models  \cite{HeydonMTWZ90, bell76,NIST_RBAC}, can be represented by CBAC policies.

Another approach consists in using term rewriting systems to express general dynamic access control policies \cite{barker06,AOliveira,KirchnerKO09,Barker-etal:07,BertolissiF14}. Term rewrite rules describe particular access control models, which can be used to verify the properties of policies by checking the confluence and termination of sets of rewrite rules. 
The approach in \cite{AlvesF15} uses rewrite rules to model the dynamics of the system and a visual graph formalism to represent a concrete state of the system.
In \cite{KirchnerKO09}, it is shown how narrowing can be used to solve administrator queries of the form “what if a request is made under these conditions?”, by representing a query as a pair of a term and a first-order equational constraint. 
Narrowing-based techniques could also be used in dynamic graph policies, since the functions defining the main relevant relations are defined by sets of rewrite rules.
The extensive theory of rewriting provides an strong platform to establish security properties \cite{AOliveira,BourdierCJK11,BertolissiF08}. Additionally, it allows using rewriting-based frameworks (such as CiME, MAUDE or TOM) to reason about those properties. The work in \cite{AlvesF15} addresses similar issues, but is based on a notion of category-based access control for distributed environments, which uses labelled graphs to include concepts like time, events, and histories. In \cite{BertolissiU13}, CiME is integrated in a tool designed to automatically check consistency and totality of RBAC access control policies. A similar technique could be used to analyse the rewrite system in a dynamic policy graph.

Several extensions of RBAC, which deal with dynamic permissions, have been proposed. These models allow permissions to change according to internal or external conditions such as time, location, or context-based properties (see, for example, \cite{ChandranJ05,JoshiBLG05,KulkarniT08}). All these extensions can be modeled as instances of CBAC.

Even though graphs are used in several models to represent and verify the properties of access control policies, there is almost no literature on tools that take advantage of the visual representation of graphs to manage the policies. The only exception that we found is the  \emph{Policy Manager} described in \cite{PolicyManager}. The \emph{Policy Manager} implements a user friendly representation of policies similar to the representation used in the \ourtool. On the other hand, in the \emph{Policy Manager}  \cite{PolicyManager}, the dynamic behaviour of CBAC policies is achieved through the edition of Ruby code, which requires the users to have knowledge of the language.

Most tools dealing with management of access control policies use tables to represent permissions data. For instance, the MotOrBAC tool \cite{Autrel_motorbac2} allows to specify and simulate policies using the ORBAC model. The ORBAC model \cite{KalamBMBCSBDT03} defines security policies centered on the organisation. Besides permissions, the model supports prohibitions and obligations. To address dynamism, the model uses \emph{contexts}, which express the conditions under which permissions are active. It includes a conflict detection feature to assist the user at finding and solving conflicts. Through its interface the user can perform all tasks related to the creation of policies and corresponding entities, which makes it very complete. However, MotOrBAC's usability has some limitations, due to the tabular representation of permissions and conflicts, which is hard to read. We believe that the usability of this kind of tool would benefit greatly from the inclusion of a graphical representation of policies.

\noindent\textit{Overview} 
 In Sec.~\ref{sec:prelim}, we recall the CBAC metamodel and the graphical framework. In Sec.~\ref{sec:architect} we describe the components of our tool. In Sec.~\ref{sec:drools} we give a brief description of Drools and illustrate how we use it to compute permissions. Finally, in Section~\ref{sec:concl} we conclude and suggest future work.

\section{Preliminaries: The Category-Based Metamodel}
\label{sec:prelim}
We assume familiarity with basic notions on first-order logic and 
term-rewriting systems~\cite{Nipkow:terraa}.
We briefly describe below the key concepts underlying the category-based 
metamodel of access control; see~\cite{Barker09} for a detailed description. We consider the following sets of entities: a countable set
$\mathcal C$ of categories, denoted $c_0$, $c_1$, $\dots$; a countable
set $\mathcal P$ of principals, denoted $p_0$, $p_1$, $\dots$; a countable set $\mathcal A$ of named
\emph{actions}, denoted $a_0$, $a_1$, $\dots$; a countable set
$\mathcal R$ of \emph{resource identifiers}, denoted $r_0$, $r_1$,
$\dots$; a finite set $\mathcal Auth$ of possible \emph{answers} to
access requests (e.g., \{\grant, \deny, {\sf undetermined}\}). 
%
The metamodel includes the following relations: 
 
\emph{Principal-category assignment:} $\mathcal P \mathcal C \mathcal
  A$ $\subseteq \mathcal P \times \mathcal C$, such that $(p,c) \in
  \mathcal P \mathcal C \mathcal A$ iff a principal $p \in \mathcal P$
  is assigned to the category $c \in \mathcal C$.

\emph{Permission-category assignment:} $\mathcal A \mathcal R \mathcal C \mathcal A$
  $\subseteq \mathcal A \times \mathcal R \times \mathcal C$, such
  that $(a,r,c) \in \mathcal A \mathcal R \mathcal C \mathcal A$ iff
  the action $a \in \mathcal A$ on resource $r \in \mathcal R$ can be
  performed by principals assigned to the category $c \in \mathcal C$.

 \emph{Authorisations:} $\mathcal P \mathcal A \mathcal R$ $\subseteq
  \mathcal P \times \mathcal A \times \mathcal R$, such that $(p,a,r)
  \in \mathcal P \mathcal A \mathcal R$ iff a principal $p \in
  \mathcal P$ can perform the action $a \in \mathcal A$ on the
  resource $r \in \mathcal R$.

Additional relations can be considered for the representation of \emph{Banned actions on resources} ($\mathcal B \mathcal A \mathcal R \mathcal C \mathcal A$) and \emph{Prohibitions} ($\mathcal B \mathcal A \mathcal R$). Throughout the paper we will use the term \emph{permissions} to refer to both \emph{Authorisations} as well as \emph{Prohibitions}, unless the distinction is required. 
 
\begin{definition}[Axioms] 
\label{def:auth} 
The relation $\mathcal P \mathcal A \mathcal R$ satisfies the
following core axiom, where we assume a relationship
$\subseteq$ between categories.
\begin{eqnarray}
  \label{eq:auth}
  \forall p \in \mathcal P,~\forall a \in \mathcal A,~\forall r \in 
\mathcal R, \quad (p,a,r)\in\mathcal{PAR} \quad\Leftrightarrow \nonumber \\ 
 \hfill \exists c, c' \in \mathcal C, ((p,c)\in\mathcal{PCA} \wedge ~ c \subseteq c' ~\wedge  (a,r,c')\in \mathcal{ARCA}) 
\end{eqnarray}
\end{definition} 
\begin{definition}[Category-based policy]
\label{def:cbpol}
A category based policy is a tuple 
$\langle\mathcal{E},\mathcal{PCA},\mathcal{ARCA},\mathcal{PAR}\rangle$, 
where $\mathcal{E} = (\mathcal{P},\mathcal{C},\mathcal{A},\mathcal{R},\mathcal{S})$, 
such that axiom (\ref{eq:auth}) is satisfied.
\end{definition} 
Axiom (\ref{eq:auth})
 states that a
request by a principal $p$ to perform the action $a$ on a resource $r$
is authorised only if $p$ belongs to a category $c$ such that for some
category $c'$ below $c$, i.e.~$c \subseteq c'$ (e.g., $c$ itself), the action
$a$ is authorised on $r$,  otherwise the request is denied.
There are other alternatives, e.g.,  considering {\em undeterminate} as answer if there is not enough information to grant the request.
Operationally, axiom (\ref{eq:auth}) can be realised through a set of
functions, as shown in~\cite{BertolissiC:essos2010}. 
 

 
\begin{definition}[Policy graph]
We define a  \emph{policy graph}, or  \emph{graph} for short, as a tuple $\mathcal{G}=(\mathcal{V},E,lv,le)$, where $\mathcal{V}$ is a set of nodes, $E$ is a set of undirected edges, which is a subset of $\{\{v_1,v_2\} \mid v_1,v_2 \in \mathcal{V} \wedge v_1 \neq v_2\}$,  $lv$ is an injective labelling function mapping nodes to entities in the category-based metamodel $lv: \mathcal{V} \rightarrow \mathcal{P} \cup \mathcal{C} \cup \mathcal{A} \cup \mathcal{R}$, and $le$ is a labelling function for edges. 
\end{definition}

We define a function $\type: \mathcal{V} \rightarrow \{P,C,A,R\}$, which associates each node with the type of its label. More precisely, $\type(v) = P$ if $lv(v)= p\in \mathcal{P}$, etc.~(that is, $P,\ C,\ A,\ R$ are the types of the nodes representing principals, categories, actions and resources, respectively). The type of an edge is determined by the type of the nodes connected by that edge, that is, an edge-type will be a pair $(T_1,T_2)$, which for simplicity we will represent as $T_1T_2$. For example, $AC$ is the type of an edge connecting a node of type $A$ with a node of type $C$. We do not distinguish between the types $T_1T_2$ and $T_2T_1$. We will use types to restrict the edges of graphs representing policies.

\begin{definition}[Well-typed policy graph] A policy graph is well typed if it contains only edges of type $PC$, $CC$, $CA$ and $AR$.
\end{definition}

\section{The \ourtool Tool}
\label{sec:architect}
The proposed tool is divided into two mains components, an engine that implements the dynamics given by the rewrite based semantics of the CBAC metamodel and a visual tool (console) that uses the representation of permissions as a graph, following~\cite{AlvesF15}, acting as an administration console for visualization and management of the access control policies.

\begin{figure}[ht]
\centering
 \includegraphics[width=1\textwidth]{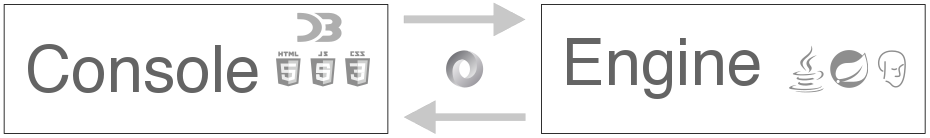}
 \caption{High level architecture}
\label{fig:DiagArchitecture}
\end{figure}

The console uses web technologies so that it can run in a web browser. The engine was developed in Java and uses the Drools rule engine to compute the permissions. Figure~\ref{fig:DiagArchitecture} resumes the technologies used in each layer.

The console was developed in HTML/Java Script and uses two main frameworks: Angular JS~\cite{AngularJS} as an MVC (Model View Controller) and IoC (Inversion of Control) platform. The D3.js framework~\cite{D3JS} is used to draw the main graph.
The server side developed in Java uses Spring (\cite{Spring}) for IoC and for exposing services through annotations. 

\subsection{Console} 

\begin{figure}[ht]
\centering
 \includegraphics[width=0.8\textwidth]{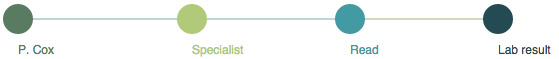}
\caption{Permission} \label{fig:LayoutGraphPar3}
\end{figure}

In the framework presented in~\cite{AlvesF15}, permissions are represented as a graph having nodes of types \textit{(P)rincipal, (C)ategory, (A)ction} and \textit{(R)esource}. A path of length three with edges \textit{PC, CA} and \textit{AR} represents a permission, i.e., the set of paths in this form represents the relation \textit{PAR}. Using colours to distinguish node types, \ourtool represents  permissions as shown as in Figure~\ref{fig:LayoutGraphPar3}.

\begin{figure}[ht]
\centering
 \includegraphics[width=0.8\textwidth]{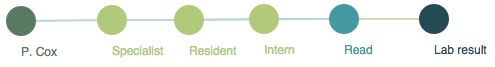}
 \caption{Inherited Permission} \label{fig:LayoutGraphPar5}
\end{figure}

The $\subseteq$ relation between categories can be represented as a directed path between two nodes CC. Therefore valid permissions can have any number of CC edges, c.f.~Figure~\ref{fig:LayoutGraphPar5}. 
This representation provides information on permissions but also on their propagation through the $\subseteq$ relation between categories, in this case, the permission is transmitted to principal P.~Cox through the chain \(Specialist \subseteq Resident \subseteq Intern\). Note that, \textit{Specialist} is a specialization of \textit{Resident}, which specializes \textit{Intern}, more broadly this relation is defined semantically as an \textit{``is a''} relation between categories which is a $\subseteq$ as required by Axiom~(\ref{eq:auth}).

\begin{figure}[ht]
\centering
 \includegraphics[width=0.8\textwidth]{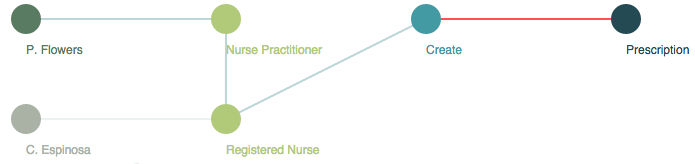}
 \caption{Inherited Prohibition} \label{fig:Layout_Graph_InherithedProhibition}
\end{figure}

Prohibitions are represented in a similar way, but the links \textit{AR} are displayed in a different colour as depicted in Figure~\ref{fig:Layout_Graph_InherithedProhibition}. 
Note that prohibitions are propagated in the inverse direction of permissions. In this case the prohibition is propagated from the Registered Nurse (or Advanced Practice Registered Nurse) to the Nurse Practicioner (the APRN is more specialized than the RN, therefore if the former is not allowed to create prescriptions then the latter is also not allowed).

A key aspect of the Category-Based Metamodel is that categories are dynamic, i.e., the assignment of principals and permissions to categories can depend on the systems state. This characteristic provides great flexibility but introduces some degree of complexity when managing and verifying permissions: 
the effective set of permissions in a certain moment depends on the systems state in that moment and on the logic associated to each of the parameters used by the permission engine; 
different system parameters can have cross effects making it difficult to predict which will be the permissions for a specific set of parameter values.

\begin{figure}[ht]
\centering
 \includegraphics[width=1.0\textwidth]{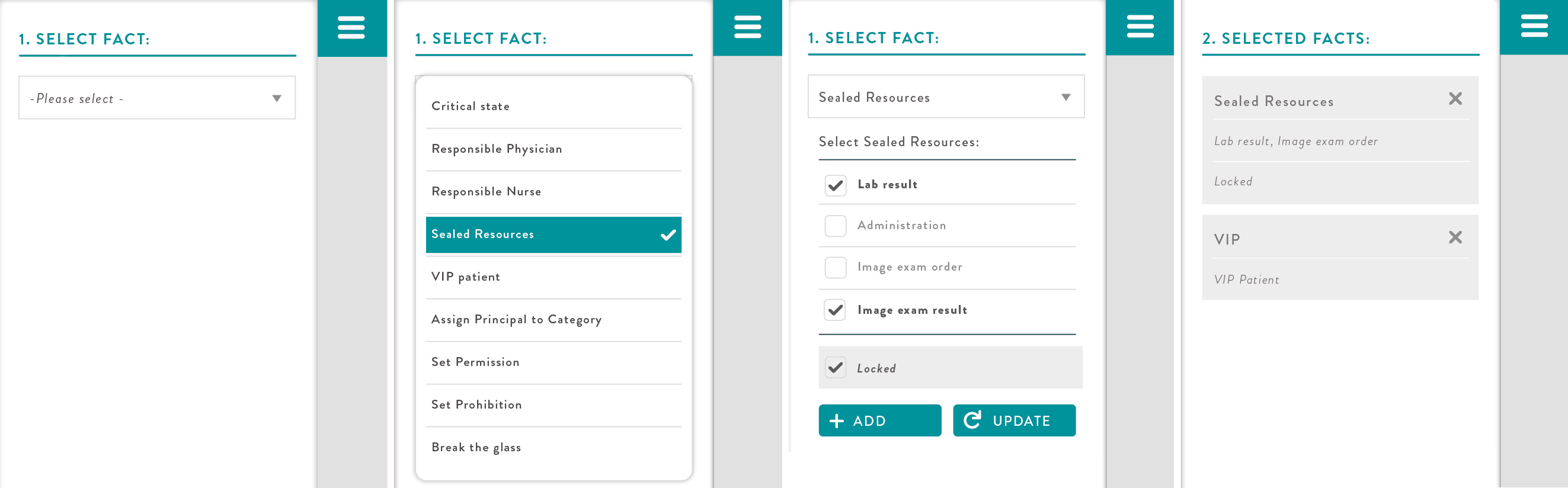}
 \caption{Custom fact selection sequence} \label{fig:Layout_Custom_FactSelectionSequence}
\end{figure}

The \ourtool console helps dealing with that complexity by providing a form of simulating scenarios for the different parameters involved in permission evaluation. As shown in Figure \ref{fig:Layout_Custom_FactSelectionSequence}, the user can choose from the available list of \textit{"Custom Facts"} and provide values for the parameters. 

The option "Update" submits the chosen facts and parameter values to the engine. The resulting graph shows the set of permissions for the system state provided, and the user can now verify the permissions for each Principal, Category, etc.

\subsection{Engine}  
The server component uses the JBoss Drools rules engine to compute permissions from a set of base configurations and user input. The normal life-cycle of the application has the following phases:

\begin{itemize}
\item~
On startup the core module loads the configuration, which includes: the set of base entities (Principal, Categories, Actions and Resources); the initial mapping Principals/Categories (\textit{Pcas}); the initial mapping between Categories and pairs Action/Resource (\textit{Arcas} for permissions and \textit{Barcas} for prohibitions); the configuration of dynamic (or customizable) facts.

Base entities are defined in a set of json files. For instance,  the list of principals is defined in the file \textit{principal.json}, which contains a list of objects of the form:

{\begin{small}
\begin{verbatim}
    {   "id": "000001",
        "name": "P. Cox",
        "title": "MD"    }
\end{verbatim}
\end{small}
}
Entity \textit{Pca} defines a mapping from a \textit{Principal} to a \textit{Category}. For example, the fact that principal with id 000001 is mapped to the category physician\textunderscore specialist, is represented by:

{\small 
\begin{verbatim}
    {   "principal": "000001",
        "category": "physician_specialist"   }
\end{verbatim}
}
Like base entities, the list of \textit{Pca} is loaded to a singleton \textit{Pcas}. In the same sense, entities \textit{Arca} (or \textit{Barca}) define which permissions (or prohibitions) are assigned to a category. 
Configuration files for these relations have similar layouts, for example, the following configuration defined as \textit{Arca} means that the category clinician can read prescriptions. The same configuration as \textit{Barca} means that the category clinician cannot read prescriptions.

{\small
\begin{verbatim}
    {   "category": "clinician",
        "action": "read",
        "resource": "prescription"  }
\end{verbatim}
}

\textit{Pcas}, \textit{Arcas} and \textit{Barcas} define a fixed set of user permissions (and prohibitions) as in a \textit{RBAC} system (more specifically a \textit{H-RBAC} since categories can be described in a hierarchical fashion).

Besides the fixed relations above, it is possible to define the set of dynamic \textit{facts} that will be available in the system.
The following excerpt contains the configuration of the fact RESPONSIBLE\textunderscore PHYSICIAN, specifying its parameters and the corresponding permitted values: 

{\small
\begin{verbatim}
    {   "fact": "RESPONSIBLE_PHYSICIAN",
        "description": "Principal is the physician responsible for 
                        the patient",
        "label": "Responsible Physician",
        "single": false, 
        "parameters": [  
           { 
             "type": "SELECTION",
             "rank": 0,
             "label": "Responsible physician",
             "description": "Responsible physician",
             "optionType": "PRINCIPAL"
           }
        ]   
    }
\end{verbatim}
}

Currently under development, is the possibility of defining in the configuration the implementation classes for Custom Facts and Parameters. That will allow extending the available facts and types of parameters by adding new implementations of the corresponding interfaces.

\item~
After initialization the services can be used by client applications to obtain the base entities, dynamic facts configuration and the possible values for the parameters of dynamic facts.

When the service that computes the permissions is called, a session of the rules engine is created. This session is fed with: a set of facts formed by the base entities, \textit{Pcas}, \textit{Arcas} and \textit{Barcas}; and a set of rules; a set of \textit{"globals"}, which are classes that provide information to compute the rules but are not themselves facts (are not used in the evaluation part of rules).  

When the session is finished the set of permissions computed is retrieved through a special class that is fed along with the set of \textit{"globals"} that is used solely to get the results. Those results are sent back to the caller.

Services are exposed in a Restful style. The complete list of services and corresponding endpoints is as follows:
\begin{itemize}
\item~List of sites, principals, categories, actions, resources: endpoints \textit{/sites}, \textit{/principals}, \textit{/categories}, \textit{/actions} and \textit{/resources}, respectively.
\item~Specific instances of each entity:  \textit{/sites/\{siteId\}}, \textit{/principals/\{principalId\}}, \textit{/categories/\{categoryId\}}, \textit{/actions/\{actionId\}} and \textit{/resources/\{resourceId\}} 
\item~Configuration of dynamic facts described above: \textit{/customFacts} 
\item~Possible values for each parameter: \textit{/customFacts/\{factId\}/params/\{rank\}/options}
\item~Permission list (accepts a list of dynamic facts values): \textit{/pars}
\end{itemize}
\end{itemize}

\section{Implementing Permissions using Drools}
\label{sec:drools}
In this section we show how permissions in our framework, defined by dynamic categories, can be computed using the Drools rule engine. We start by giving a brief description of the Drools Rule Engine (see~\cite{drools} for full documentation). 
\subsection{Drools}
The Drools rule engine is based on the Rete pattern matching algorithm~\cite{Forgy1982} for production rule systems, which implements forward chaining using directed acyclic graphs to represent rules (naive implementation of forward chaining does not scale well with an increasing number of rules). We start by describing the Drools native rule language. Very briefly, a rule is a declaration of the form:

{\small
\begin{verbatim}
  rule "name"
     Attributes
  when     
     LHS
  then
     RHS
  end
\end{verbatim}
}
where {\tt name} is the unique identifier of the rule, {\tt Attributes} is a list of optional features that can influence the behaviour of the rule, {\tt LHS} (Left Hand Side) specifies a particular set of conditions, and {\tt RHS} (Right Hand Side) is a block of code specifying the actions to be executed, should the conditions in {\tt LHS} be satisfied. 

The list {\tt Attributes} is a (possibly empty) sequence of  attributes from the following set: 
$\{${\tt no-loop}, {\tt ruleflow-group}, {\tt lock-on-active}, {\tt salience}, {\tt dialect}, {\tt agenda-group}, {\tt auto-focus}, {\tt activation-group},\ \\ {\tt date-effective}, {\tt date-expires}, {\tt duration}$\}$.
\ The attribute {\tt salience} allows to define priority between rules. The default value for salience is zero, but rules with higher salience have priority over rules with lower salience. This attribute can also be defined dynamically using bound variables. 

{\tt LHS} consists of zero or more Conditional Elements, which determines when the rule applies. If {\tt LHS} is empty, then the conditional element is considered to always be true, causing the rule to be activated once when a new session is created (in Drools rules are evaluated in the context of a \textit{Session} into which data can be inserted and from which process instances can be created). Conditional elements work on patterns, which can match on facts currently in the working memory. Patterns can contain zero or more constraints. Because of the amplitude of options, we do not describe the full syntax for conditional elements. We refer the reader to~\cite{drools}, for the full description. 

The {\tt RHS} represents the action part of the rule, whose main objective is to insert, modify or delete data in the working memory. This part of the rule corresponds to code in a supported dialect. It should be atomic and not contain conditional/imperative code, since it represents what to do when {\tt LHS} is valid. There are methods available for conveniently changing facts in memory, such as: {\tt update(object,handle)}, {\tt update(object)}, {\tt insert(new Object())}, {\tt delete(handle)}, etc. Again, we refer to~\cite{drools}, for all available methods.

As stated before, rules are evaluated within a session, which can be of two types: \textit{Stateless} or \textit{Stateful}. \ourtool uses a \textit{Stateful} session to evaluate rules because this kind of session allows inference, i.e., if rules change data in the session then rules are reevaluated which can lead to new rules being triggered. In both cases, the session is fed with the rules and two other kinds of data:
\paragraph{Globals: }  
Globals are objects that are made visible to the rule engine but changes in them do not trigger reevaluation of rules.  They provide context information that can be used to evaluate rules and as a mean of returning objects from the engine. In \ourtool, the class \textit{Pars()} is used as a global to get the resulting permissions.
\paragraph{Facts: } 
\ourtool's ontology is composed by the base entities, \textit{Principal}, \textit{Category}, \textit{Action} and \textit{Resource}, by the fixed relations between those entities, \textit{Pca}, \textit{Arca} and \textit{Barca} and by the set of customizable entities. 
\ The data used by the rules are instances of those objects, also called \textit{facts} in the Drools language. In the \ourtool implementation, all those classes implement the same interface.

When facts in the session satisfy the condition in the {\tt LHS} of a rule, then the actions specified in the consequence ({\tt RHS}) are triggered. For example, \ourtool allows mapping explicitly a principal to a category.

{\small
\begin{verbatim}
rule "Rule add customs Pcas" 
    when
     $pca : SetPca() 
    then 
     insert(new Pca($pca.getPrincipal(), $pca.getCategory())); 
end
\end{verbatim}
}
The only restriction on the above condition is the existence of a \textit{SetPca} fact. That means that, for each \textit{SetPca} fact in the session, a new \textit{Pca} fact will be created and that \textit{Pca} will have the same value for its attributes \textit{Principal} and \textit{Category} as the fact who triggered its creation.

\subsection{Rule Processing} 

File \textit{custom.drl }contains all rules that process the customizable facts. All these rules modify, insert or delete \textit{Pca}, \textit{Arca} and \textit{Barca} facts. The final set of permissions is then computed by the following pair of rules:

{\small
\begin{verbatim}
rule "Pars - Permissions"
  salience -100 
  when
    $principal : Principal( $pid : id )
    $category : Category( $cid : id )
    $pca : Pca(principal.id == $pid, category.id ==  $cid)
    $arca : Arca(categories.containsOrEquals(category.id, $cid))  	
  then    
    pars.add(
      new Par(
        $principal, 
        categories.getPermissionChain($cid, $arca.category.id), 
        $arca.permission
      ) )    		    
end

rule "Pars - Prohibitions"
  salience -100 
  when
    $principal : Principal( $pid : id )
    $category : Category( $cid : id )
    $pca : Pca(principal.id == $pid, category.id ==  $cid)
    $barca : Barca(categories.containsOrEquals($cid, category.id))	
  then
    pars.add(
      new Par(
        $principal, 
        categories.getProhibitionChain($cid, $barca.category.id), 
        $barca.permission
      ) )    		    
end
\end{verbatim}
}
The first rule computes the permissions and the second computes the prohibitions. In both cases, the condition finds all \textit{Pca} and \textit{Arca} with matching categories (like a database join). The consequence inserts a new \textit{Par} in \textit{Pars} for each match (\textit{Pars} is the provided instance of \textit{Pars()} that allows retrieving the results from the engine).

Function \textit{categories.containsOrEquals(a,b)} returns true if \textit{a} and \textit{b} are the same or if \textit{a} is a superset of \textit{b}.
This function provides the hierarchical logic applied to categories. Parent categories are more general than their children (defining an \textit{``is a''} relation between child and parent), this means that if a parent has a permission then its children inherits it. Prohibitions, on the other hand, are propagated from more specific to more general categories, that is why the function \textit{containsOrEquals()}, in rule "Prohibitions", has the parameters switched (equivalent to a function \textit{isContainedOrEquals()}).

Functions \textit{getPermissionChain(a,b)} and \textit{getProhibitionChain(a,b)} make sure that \textit{Pars} includes the hierarchy for permissions that are propagated. For instance, a \textit{Par} of the form \textit{\{P, [C1, C2], Perm\}} means that principal \textit{P} is mapped to category \textit{C1}, which has permission \textit{Perm} but that permission was inherited from \textit{C2}. Figure \ref{fig:Layout_Graph_InherithedProhibition} shows how the hierarchy information is displayed in the console.



The two rules above implement  almost directly the operational realisation of axiom (\ref{eq:auth}), given by the following rewriting rule, cf.~\cite{BertolissiC:essos2010}:

{\small
\[
\begin{array}{ll}
par(p,a,r) & \rightarrow\ if(a, r) \in arca^* (below(pca(p)))\ then\ grant \\
           & else\ if(a, r) \in barca^*(above(pca(p)))\ then\ deny\ else\ undeterminate
\end{array}
\]
}

There is, however, a subtle but important difference between the axiom and rules above. The \textit{if} clause in the axiom implies that there are no conflicting permissions in the result, i.e a category can not be simultaneously mapped to a permission  and to a prohibition on the same action/resource. To solve this issue an additional rule is needed:

{\small
\begin{verbatim}
rule "Pars - Conflicts - Remove Barca"
	salience -60
    when
    	Category($cId : id)
    	Action($aId : id)
    	Resource($rId : id)
    	$barca : Barca(
    	  categories.containsOrEquals($cId, category.id), 
    	  permission.action.id == $aId, 
    	  permission.resource.id == $rId)
    	Arca(
    	  category.id == $cId, 
    	  permission.action.id == $aId, 
    	  permission.resource.id == $rId)
    then
		delete($barca)
end
\end{verbatim}
}

This rule removes the \textit{Barca} if there is an \textit{Arca} for the same pair action/resource and category (or some category that generalizes it). This means that the system prioritizes permissions over prohibitions. To prioritize prohibitions this rule should be replaced by an equivalent that removes the conflicting \textit{Arca} instead.
By associating the value of {\tt salience} to a bound variable that can be customised by the user, priority between permissions and prohibitions can be changed dynamically.
\subsection{Rules for Customizable Facts} 
Customizable facts are used to simulate system conditions that affect users permissions. The service responsible for the evaluation of permissions accepts a list of facts that are injected in the Drools session. Each custom fact has at least one rule that implements its logic, as illustrated by the following examples.

\paragraph{Scenario One.} Suppose that in a hospital clinical records are normally accessible only to the patient's responsible physician and nurse, but, if the patient is in critical state, then all clinicians have access to the patient's record.
A simple way to achieve this is verifying that, if critical status is set then all clinicians (nurses or physicians) are mapped to a special category that has reading permissions on the record. A rule to process this logic is as follows:

{\small
\begin{verbatim}
rule "Rule critical state - read all"
  when
    CriticalState(criticalState == Boolean.TRUE)
    $principal : Principal()
    Category($cid : id, id == "clinician")
    $pca : Pca(principal.id == $principal.id, 
      categories.containsOrEquals($cid, category.id))
  then
    insert(new Pca($principal, 
                   categories.getCategoryById("read_all")));    		    
end
\end{verbatim}
}

This rule maps every principal that is associated to a specialization of \textit{clinician} to the category \textit{read\_all} (i.e.,~inserts a \textit{Pca} for each principal with category \textit{read\_all} ).
Additionally, if the rule \textit{"Pars - Conflicts"} is configured to prioritize prohibitions over permissions, then  an extra rule is needed to remove eventual reading prohibitions so that the resulting pars are as expected: 

{\small
\begin{verbatim}
rule "Rule critical state - removed read prohibitions"
  when
    CriticalState( criticalState == Boolean.TRUE )
    Category( $cid : id, id == "clinician" )
    $barca : Barca(
      categories.containsOrEquals($cid, category.id), 
      permission.action.id == "read"
    )  	 
  then
    delete($barca)   
end
\end{verbatim}
}
This rule removes all \textit{Barcas} for action \textit{read} that have a category that is a specialization of \textit{clinician}, i.e., all reading prohibitions for clinicians are removed.

\paragraph{Scenario Two.} In some countries, patients can request restrictions on their clinical records, for instance in the UK a patient can \textit{"Seal"} or \textit{"Seal and Lock"} parts of its clinical record (refer to ~\cite{HSCIC:PatientChoices} for more details). 
If a patient request to \textit{"Seal and Lock"} a part of its record then nobody can access it no matter the circumstances. That result can be achieved by a rule that removes all permissions for locked resources:

{\small
\begin{verbatim}
rule "Sealed and Locked resources"
  when
    SealedResource ($resource: resource, locked == Boolean.TRUE)
    $arca : Arca(permission.resource.id == $resource.id)
  then
    delete($arca);
end
\end{verbatim}
}
On the other hand, if a patient asks to \textit{"Seal"} its record then data will be hidden by default but clinicians can break that seal (in which case that action will be logged and security managers notified). That process is also called \textit{"Break the Glass"}~\cite{Povey:1999}.
The rule to implement that logic is more complex than the previous one because it has to make sure that only the principal who performed the \textit{"Break the Glass"} gets access to the record. The following rule describes one possible way of doing it:

{\small
\begin{verbatim}
rule "Sealed resources"
  when
    $catSealed : Category(id == "sealed_resource")
    SealedResource (
      $resource : resource, 
      locked == Boolean.FALSE      
    )
    $arca : Arca(
      category.id != $catSealed.id, 
      permission.resource.id == $resource.id
    ) 
    $pca : Pca(
      $principal : principal, 
      category == $arca.category
    )
    BreakTheGlass(principal.id == $principal.id)
  then
    Permission permission = 
      (Permission) PermissionFactory.buildPermission(
        $arca.permission.getAction(), 
        $resource
      )
    delete($arca)
    insert(new Arca($catSealed, permission))
    insert(new Pca($principal, $catSealed))
end
\end{verbatim}
}
This rules uses an auxiliary category \textit{"sealed\_resources"}. All \textit{Arca} on sealed resources are removed from its original categories and mapped to this category. Then, every \textit{Principal} that performed \textit{"Break the Glass"} and was mapped to the original category is also mapped to the category \textit{"sealed\_resources"}.

Note that, even though all these rules apply very different logics, their consequences always create, modify or update \textit{Pcas}, \textit{Arcas} and \textit{Barcas}, i.e., these rules affect the mapping between principals and categories and between categories and permissions or prohibitions. The result of processing these rules is a set of facts that reflects the base set and all the interactions introduced by the customizable facts. That set is then processed by the rules that compute the resulting \textit{Pars} (Rules \textit{Pars - Permissions}, \textit{Pars - Prohibitions} and \textit{Pars - Conflicts}).

\section{Conclusions and Future Work}
\label{sec:concl}
We present a novel approach to access control using the Drools rule engine to implement the dynamic aspects of access control policies. Rules are used both to compute permissions from given relations (between users and categories, and categories and actions on resources), 
 but also permissions that depend on customizable facts of the system. Together with a visual tool, this provides an efficient way to explore how changes in the system can affect permission assignment. Additional functionalities to improve permission analysis are under development. A suitable display of the changes in the policy graph, resulting from two different sets of facts/parameters.  An option allowing the system manager to filter particular principals, actions and resources and solely display the associated policy (sub-)graph. Representation of principals, categories, actions or resources, connected to the same nodes, by compound nodes, which can be unfolded whenever required. This results in more compact and intelligible graphs, also allowing identification of entities with the same properties.
\bibliographystyle{abbrv}
\bibliography{main}

\begin{thebibliography}{10}

\bibitem{AliF14}
A.~Ali and M.~Fern{\'{a}}ndez.
\newblock Hybrid enforcement of category-based access control.
\newblock In {\em Security and Trust Management - 10th International Workshop,
  {STM} 2014, Wroclaw, Poland, September 10-11, 2014. Proceedings}, pages
  178--182, 2014.

\bibitem{AlvesF15}
S.~Alves and M.~Fern{\'{a}}ndez.
\newblock A framework for the analysis of access control policies with
  emergency management.
\newblock {\em ENTCS}, 312:89--105, 2015.

\bibitem{Autrel_motorbac2}
F.~Autrel, F.~Cuppens, N.~Cuppens-Boulahia, and C.~Coma-Brebel.
\newblock {MotOrBAC 2: a security policy tool}.
\newblock In {\em SARSSI'08 : 3{\`e}me conf{\'e}rence sur la S{\'e}curit{\'e}
  des Architectures R{\'e}seaux et des Syst{\`e}mes d'Information, 13-17
  octobre, Loctudy, France}, 2008.

\bibitem{Nipkow:terraa}
F.~Baader and T.~Nipkow.
\newblock {\em Term rewriting and all that}.
\newblock Cambridge University Press, Great Britain, 1998.

\bibitem{Barker09}
S.~Barker.
\newblock {The Next 700 Access Control Models or a Unifying Meta-model?}
\newblock In {\em Proceedings of the 14th ACM Symposium on Access Control
  Models and Technologies}, SACMAT '09, pages 187--196, New York, NY, USA,
  2009. ACM.

\bibitem{barker06}
S.~Barker and M.~Fern{\'{a}}ndez.
\newblock Term rewriting for access control.
\newblock In {\em Data and Applications Security XX, 20th Annual {IFIP} {WG}
  11.3 Working Conference on Data and Applications Security, Sophia Antipolis,
  France, July 31-August 2, 2006, Proceedings}, pages 179--193, 2006.

\bibitem{bell76}
D.~E. Bell and L.~J. LaPadula.
\newblock Secure computer system: Unified exposition and multics
  interpretation.
\newblock {\em MITRE-2997}, 1976.

\bibitem{BertolissiF08}
C.~Bertolissi and M.~Fern{\'{a}}ndez.
\newblock A rewriting framework for the composition of access control policies.
\newblock In {\em Proceedings of the 10th International {ACM} {SIGPLAN}
  Conference on Principles and Practice of Declarative Programming, July 15-17,
  2008, Valencia, Spain}, pages 217--225, 2008.

\bibitem{BertolissiC:essos2010}
C.~Bertolissi and M.~Fern{\'a}ndez.
\newblock Category-based authorisation models: operational semantics and
  expressive power.
\newblock In {\em Proc. of ESSOS 2010}, volume 5965 LNCS of {\em LNCS}, pages
  283 -- 301. Springer, 2010.

\bibitem{BertolissiC:STM2010}
C.~Bertolissi and M.~Fern{\'a}ndez.
\newblock Rewrite specifications of access control policies in distributed
  environments.
\newblock In {\em Proc. of STM 2010: 6th Workshop on Security and Trust
  Management, Athens, G reece, 2010}, number 6710 in Lecture Notes in Computer
  Science. Springer, 2011.

\bibitem{BertolissiF14}
C.~Bertolissi and M.~Fern{\'{a}}ndez.
\newblock A metamodel of access control for distributed environments:
  Applications and properties.
\newblock {\em Inf. Comput.}, 238:187--207, 2014.

\bibitem{Barker-etal:07}
C.~Bertolissi, M.~Fern{\'a}ndez, and S.~Barker.
\newblock Dynamic event-based access control as term rewriting.
\newblock In {\em DBSec}, pages 195--210, 2007.

\bibitem{BertolissiU13}
C.~Bertolissi and W.~Uttha.
\newblock Automated analysis of rule-based access control policies.
\newblock In {\em Proc. of PLPV}, pages 47--56, 2013.

\bibitem{D3JS}
M.~Bostock.
\newblock {D3.js}.
\newblock \url{http://d3js.org/}, 2015.
\newblock [Online; accessed 01-June-2015].

\bibitem{BourdierCJK11}
T.~Bourdier, H.~Cirstea, M.~Jaume, and H.~Kirchner.
\newblock Formal specification and validation of security policies.
\newblock In {\em Foundations and Practice of Security - 4th Canada-France
  {MITACS} Workshop, {FPS} 2011, Paris, France, May 12-13, 2011, Revised
  Selected Papers}, pages 148--163, 2011.

\bibitem{ChandranJ05}
S.~M. Chandran and J.~B.~D. Joshi.
\newblock \emph{LoT-RBAC}: {A} location and time-based {RBAC} model.
\newblock In {\em Web Information Systems Engineering - {WISE} 2005, 6th
  International Conference on Web Information Systems Engineering, New York,
  NY, USA, November 20-22, 2005, Proceedings}, pages 361--375, 2005.

\bibitem{AOliveira}
A.~S. de~Oliveira.
\newblock {\em Réécriture et Modularité pour les Politiques de Sécurité}.
\newblock PhD thesis, Université Henri Poincaré, Nancy, France, 2008.

\bibitem{Forgy1982}
C.~L. Forgy.
\newblock Rete: A fast algorithm for the many pattern/many object pattern match
  problem.
\newblock {\em Artificial Intelligence}, 19(1):17 -- 37, 1982.

\bibitem{AngularJS}
Google.
\newblock {AngularJS}.
\newblock \url{https://angularjs.org/}, 2015.
\newblock [Online; accessed 01-June-2015].

\bibitem{drools}
R.~Hat.
\newblock {Drools}.
\newblock \url{http://www.drools.org}, 2015.
\newblock [Online; accessed 27-May-2015].

\bibitem{HSCIC:PatientChoices}
Health and P.~C.~I. Center.
\newblock {Patient choices}.
\newblock \url{http://systems.hscic.gov.uk/infogov/confidentiality/choices},
  2015.
\newblock [Online; accessed 2-March-2015].

\bibitem{HeydonMTWZ90}
A.~Heydon, M.~W. Maimone, J.~D. Tygar, J.~M. Wing, and A.~M. Zaremski.
\newblock {Mir{\'{o}}: Visual Specification of Security}.
\newblock {\em {IEEE} Trans. Software Eng.}, 16(10):1185--1197, 1990.

\bibitem{Hoagland2000}
J.~A. Hoagland.
\newblock {Specifying and Implementing Security Policies Using LaSCO, the
  Language for Security Constraints on Objects}.
\newblock {\em CoRR}, cs.CR/0003066, 2000.

\bibitem{JoshiBLG05}
J.~Joshi, E.~Bertino, U.~Latif, and A.~Ghafoor.
\newblock A generalized temporal role-based access control model.
\newblock {\em {IEEE} Trans. Knowl. Data Eng.}, 17(1):4--23, 2005.

\bibitem{KalamBMBCSBDT03}
A.~A.~E. Kalam, S.~Benferhat, A.~Mi{\`{e}}ge, R.~E. Baida, F.~Cuppens,
  C.~Saurel, P.~Balbiani, Y.~Deswarte, and G.~Trouessin.
\newblock {Organization based access contro}.
\newblock In {\em 4th {IEEE} International Workshop on Policies for Distributed
  Systems and Networks {(POLICY} 2003), 4-6 June 2003, Lake Como, Italy}, page
  120, 2003.

\bibitem{KirchnerKO09}
C.~Kirchner, H.~Kirchner, and A.~S. de~Oliveira.
\newblock Analysis of rewrite-based access control policies.
\newblock {\em Electr. Notes Theor. Comput. Sci.}, 234:55--75, 2009.

\bibitem{KochMP02}
M.~Koch, L.~V. Mancini, and F.~Parisi{-}Presicce.
\newblock {A graph-based formalism for RBAC}.
\newblock {\em {ACM} Trans. Inf. Syst. Secur.}, 5(3):332--365, 2002.

\bibitem{KochMP05}
M.~Koch, L.~V. Mancini, and F.~Parisi{-}Presicce.
\newblock {Graph-based specification of access control policies}.
\newblock {\em J. Comput. Syst. Sci.}, 71(1):1--33, 2005.

\bibitem{KulkarniT08}
D.~Kulkarni and A.~Tripathi.
\newblock Context-aware role-based access control in pervasive computing
  systems.
\newblock In {\em {SACMAT} 2008, 13th {ACM} Symposium on Access Control Models
  and Technologies, Estes Park, CO, USA, June 11-13, 2008, Proceedings}, pages
  113--122, 2008.

\bibitem{PolicyManager}
H.~Mirzapour-Aghdaghi and M.~Fernández.
\newblock {Policy Manager: a tool to analyse category-based access control
  policies}.
\newblock \url{http://policymanager.herokuapp.com}, 2014.
\newblock [Online; accessed 24-August-2015].

\bibitem{Spring}
I.~Pivotal~Software.
\newblock {Spring}.
\newblock \url{https://spring.io/}, 2015.
\newblock [Online; accessed 01-June-2015].

\bibitem{Povey:1999}
D.~Povey.
\newblock Optimistic security: A new access control paradigm.
\newblock In {\em Proceedings of Workshop on New Security Paradigms}, NSPW '99,
  pages 40--45. ACM, 2000.

\bibitem{NIST_RBAC}
R.~K. Ravi~Sandhu, David~Ferraiolo.
\newblock {The NIST Model for Role-Based Access Control: Towards A Unified
  Standard}.
\newblock In {\em Proceedings, 5th ACM Workshop on Role Based Access Control},
  pages 47--63, 2000.

\bibitem{sandhu96}
R.~Sandhu, E.~Coyne, H.~Feinstein, and C.~Youman.
\newblock Role-based access control models.
\newblock {\em IEEE Computer}, 29(2):38--47, 1996.

\end{thebibliography}

\end{document}